\documentclass[a4paper,fleqn]{cas-dc}

\usepackage[authoryear,longnamesfirst]{natbib}

\def\tsc#1{\csdef{#1}{\textsc{\lowercase{#1}}\xspace}}
\tsc{WGM}
\tsc{QE}
\begin{document}
\let\WriteBookmarks\relax
\def\floatpagepagefraction{1}
\def\textpagefraction{.001}

\title [mode = title]{Single-run determination of the saturation vapor pressure and enthalpy of vaporization/sublimation of a substance undergoing successive solid-solid and solid-liquid phase transitions: the case of $N$-methyl acetamide}

\affiliation{organization={Department of Physics and Astronomy},
            addressline={Aarhus University}, 
            city={Aarhus},
          citysep={}, % Uncomment if no comma needed between city and postcode
            postcode={DK 8000}, 
            %state={},
            country={Denmark}}

\author{Mohsen Salimi}
\credit{}

\author{Aur\'{e}lien Dantan}
\credit{}

\author[1]{Henrik B. Pedersen}

\cormark[1]
\cortext[1]{Corresponding author: H. B. Pedersen, hbp@phys.au.dk}

\credit{}


\begin{abstract}
We report on the dynamical measurement of the saturation vapor pressure of $N$-methyl acetamide in the temperature range $-30^\circ$C to $34^\circ$C. This is achieved by monitoring the pressure inside a vacuum chamber in which a precooled sample of the substance slowly thermalizes to the chamber temperature, undergoing first a phase transition between two crystalline structures around $1^\circ$C and then a solid-liquid phase transition around $30^\circ$C. Such a measurement provides in a single run accurate data for the saturation vapor pressure and the enthalpies of sublimation and vaporization of the different phases of the investigated substance.
\end{abstract}

\begin{highlights}
\item Single run-measurement of the saturation vapor pressure of a polymorphic substance undergoing solid-solid and solid-liquid phase transitions over an extended temperature range around room temperature
\item Accurate thermodynamics data for $N$-methyl acetamide in its crI, crII and liquid phases in the range $-30^\circ$C to $34^\circ$C
\item First determination of the saturation vapor pressure and enthalpy of sublimation for crII $N$-methyl acetamide
\end{highlights}


\maketitle

% Main text
\section{Introduction}
\label{sec:introduction}

$N$-methyl acetamide (CH$_3$NHCOCH$_3$)---the $N$-methyl derivative of acetamide---benefits from a high dipole moment, low mass and low volatility and being amphiphilic. As such, it is is widely used as a solvent~\cite{Ashford1995}, in the manufacturing of pharmaceuticals, pesticides, polymers and batteries~\cite{Bernauer2008} and in electrochemistry applications~\cite{Kirk1991}. Due to the peptide bond, acetamides are also relevant possible interstellar indicators of the existence of life~\cite{Halfen2011}.

$N$-methyl acetamide (NMA) is solid at room temperature and has a melting phase transition temperature of $\sim31^\circ$C. Interestingly, it is a polymorphic substance with two eniantropic structures~\cite{Katz1960} and a phase transition (when heated) from crII to crI at $\sim1^\circ$C.

The thermodynamics properties of $N$-methyl acetamide around room temperature have been the focus of recent studies~\cite{Zaitseva2019,Stejfa2020}, in which accurate data for the saturation vapor pressure (SVP) and the enthalpy of vaporization for liquid NMA and the enthalpy of sublimation for crI NMA have been reported. 

In this work we make use of a newly established dynamical method~\cite{Nielsen2024,Salimi2025b} to accurately determine the SVP of a low-volatile substance by isolating it in a static vacuum chamber and monitoring the chamber pressure as the precooled substance slowly thermalizes to the higher chamber temperature. We apply this method to $N$-methyl acetamide and show that it allows---in a single run---to accurately determine its SVP, and subsequently the enthalpies of sublimation/vaporization of the different phases, as it successively undergoes a solid-solid and a solid-liquid phase transition. This allows in particular to provide for the first time SVP and enthalpy of sublimation data for crII NMA in the range $-30$ to $0^\circ$C.

%The paper is organized as follows: the measurement procedure, the purification sequence and the theoretical model used for the analysis of the experimental data are introduced in Sec.~\ref{sec:methods}. The experimental results and reported thermodynamical data are presented in Section~\ref{sec:results} and Section~\ref{sec:conclusion} concludes.

%%%%%%%%%%%%%%%%%%%%%%%

\section{Methods}
\label{sec:methods}

\subsection{Measurement procedure}

The ASVAP experimental apparatus has been described in detail in Refs.~\cite{Nielsen2024,Salimi2025b}. It consists in three interconnected vacuum chambers: {\it (i)} a load chamber, in which the sample is inserted at room temperature and purified under low- and high-vacuum conditions, as will be detailed in the next section, {\it (ii)} a transfer chamber, in which the sample is precooled to a temperature well below room temperature ($-30^\circ$C in the experiments reported here) under high vacuum conditions ($\sim 10^{-6}$ Pa), and {\it (iii)} an experimental chamber with a temperature stabilized at or above room temperature and in which a high vacuum ($<10^{-6}$ Pa) has been realized prior to the insertion of the sample. The pressure in the experimental chamber is then monitored by means of an absolute pressure sensor (capacitive diaphragm BCEL7045 0.1 mbar, Edwards) under static vacuum conditions (outgassing rate of the chamber$\sim10^{-2}$ Pa/h at 35$^\circ$C) as the sample slowly thermalizes to the chamber temperature (thermalization time$\sim$1 h). 

\begin{figure}[h]
  \centering
   \includegraphics[width=\columnwidth]{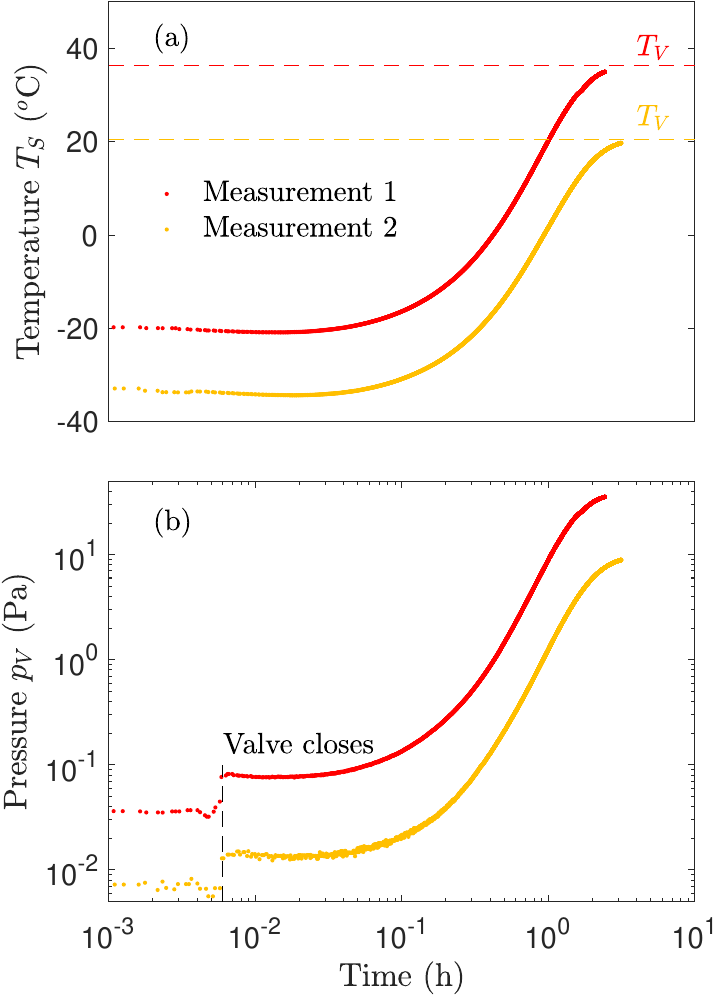}
    \caption{$T_s$ and $p_V$ as a function of time for {\it measurement 1} (red, initial sample temperature of $-20^\circ$C and chamber temperature of 34.5$^\circ$C) and {\it measurement 2} (orange, initial sample temperature of $-34^\circ$C and chamber temperature of 20.5$^\circ$C).}\label{fig:time}
\end{figure}

Two examples of the measured sample temperature and chamber pressure are shown in Fig.~\ref{fig:time}: {\it measurement 1} for an initial sample temperature of $-20^\circ$C and a chamber temperature of 34.5$^\circ$C and {\it measurement 2} for an initial sample temperature of $-34^\circ$C and a chamber temperature of 20.5$^\circ$C. Exponential fits of the temperature variations in time yield time constants of $\sim 3000$ s.

\subsection{Purification}

\begin{figure}[h]
  \centering
    \includegraphics[width=\columnwidth]{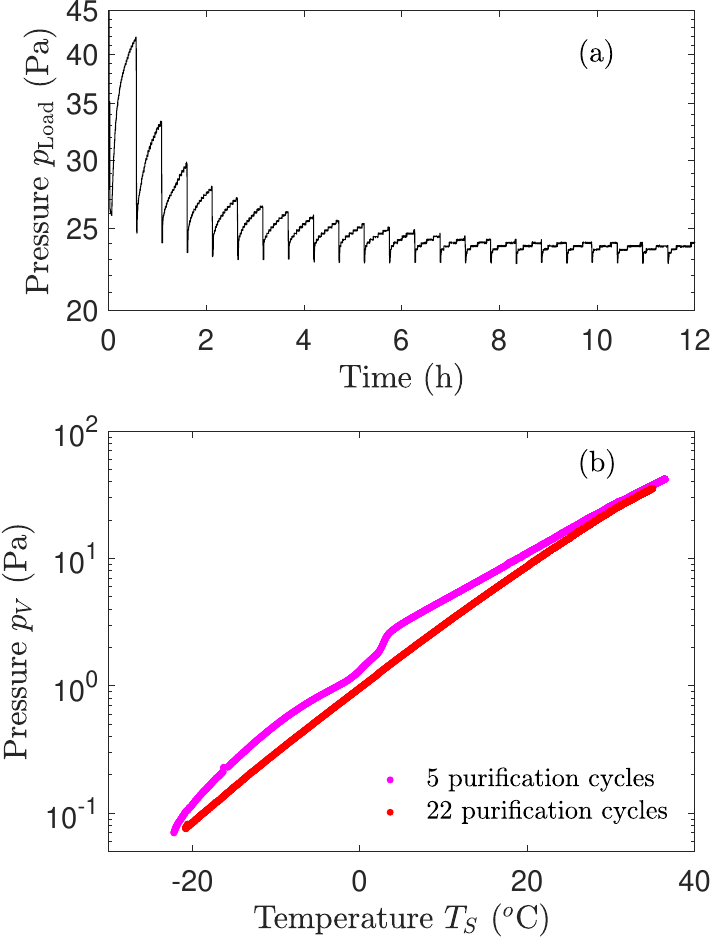}
    \caption{(a) Example of variation of the pressure in the load chamber under successive cycles of evacuation and static vacuum. (b) Examples of $p_V$ {\it vs} $T_S$ measurements after short (5 cycles, magenta) and long (22 cycles, red) purification sequences in the conditions corresponding to those of measurement 1.}\label{fig:purification}
\end{figure}

The $N$-methyl acetamide sample was provided by Sigma-Aldrich/Merck with a certified purity of XXX\%. Prior to insertion in the load chamber the sample is heated slightly above the melting temperature and 2-3 mL are deposited in the sample holder in the load chamber slightly overpressurized with nitrogen. However, due to the relatively high hydrophilic nature of $N$-methyl acetamide, the sample needs to be further purified after insertion. This is realized by performing repeated cycles of evacuation (active pumping with a scroll turbo pump and evaporative cooling) and reheating under static vacuum conditions (although in presence of an active vacuum gauge), whereby substances with a higher saturation vapor pressure (e.g. water) are gradually removed from the sample. A typical variation of pressure in the load chamber during purification is shown in Fig.~\ref{fig:purification}(a), where the peak pressure in the reheating phase is observed to decrease with the number of cycles and converges towards a constant value. Subsequent measurements of the experimental chamber pressure $p_V$ versus the sample temperature $T_S$ during thermalization with the experimental chamber temperature show that insufficient purification leads to an overestimation of the saturation vapor pressure, as is illustrated in Fig.~\ref{fig:purification}(b). The crII-crI phase transition around $1^\circ$C is in particular clearly visible for the $p_V$ {\it vs} $T_S$ curve corresponding to a short purification sequence (blue), which we surmize is due to the sudden release of water from the sample as the phase transition occurs. For a sufficiently long purification, though, and as long as the amount of sample remaining prior to the measurement in the experimental chamber is not vanishingly small, the experimentally measured $p_V$ {\it vs} $T_s$ curves reproducibly converge to the one represented by the red curve.

\subsection{Determination of the saturation vapor pressure}

To determine the SVPs and enthalpies of $N$-methyl acetamide in its different phases as a function of temperature, we follow the method described in Ref.~\cite{Salimi2025b}. 

For an ideal gas, the temperature dependence of the SVP is given by the Clausius-Clapeyron equation~\cite{wark1988}
\begin{equation}
\frac{dp_\textrm{sat,u}}{dT}=\frac{\Delta H_\textrm{u}}{k_\textrm{B}N_\textrm{A}T^2}\times p_\textrm{sat,u},
\label{eq:dpsat}
\end{equation}
where $p_\textrm{sat,u}$ and $\Delta H_\textrm{u}$ are, respectively, the SVP and the enthalpy of vaporization/sublimation for the phase considered ($\textrm{u}=\textrm{I},\textrm{II},\ell$), $k_\textrm{B}$ is Boltzmann's constant and $N_\textrm{A}$ is Avogadro's number. Since we are interested in taking into account the variation of the enthalpy of sublimation/vaporization with temperature in a limited temperature range around a given temperature $T^*_u$. These variations are related to the difference in the heat capacities at constant pressure of the gas and solid/liquid phases, $C_{p,\textrm{g}}$ and $C_{p,\textrm{u}}$, by
\begin{equation}
\frac{d\Delta H_\textrm{u}}{dT}=C_{p,\textrm{g}}-C_{p,\textrm{u}},
\end{equation}
which can be well-approximated in the ranges considered by linear variations around $T^*_\textrm{u}$~\cite{Clarke1966}
\begin{equation}C_{p,\textrm{g}}-C_{p,\textrm{u}}\simeq \beta_\textrm{u}+\alpha_\textrm{u}(T-T^*_\textrm{u}),
\end{equation}
where $\beta_\textrm{u}$ and $\alpha_\textrm{u}$ are constants. Integrating this equation yields
\begin{equation}
\Delta H_\textrm{u}=\Delta H_\textrm{u}^*+\beta_\textrm{u}(T-T^*_\textrm{u})+\frac{1}{2}\alpha_\textrm{u}(T-T^*_\textrm{u})^2,
\label{eq:DH}
\end{equation}
where $\Delta H_\textrm{u}^*$ is the enthalpy of vaporization/sublimation at $T^*_\textrm{u}$. Integrating Eq.~(\ref{eq:dpsat}) with the expression for $\Delta H_\textrm{u}$ given by Eq.~(\ref{eq:DH}) yields an expected variation of the saturation vapor pressure with temperature given by
\begin{align}
\nonumber p_\textrm{sat,u}=&p^*_\textrm{sat,u}\exp\left[-\frac{\Delta H_\textrm{u}^*}{k_\textrm{B}N_\textrm{A}}\left(\frac{1}{T}-\frac{1}{T^*_\textrm{u}}\right)\right.\\
\nonumber &-\frac{\beta_\textrm{u}}{k_\textrm{B}N_\textrm{A}}\left(1-\frac{T^*_\textrm{u}}{T}+\ln\frac{T^*_\textrm{u}}{T}\right)\\
&\left. +\frac{\alpha_\textrm{u} T^*_\textrm{u}}{2k_\textrm{B}N_\textrm{A}}\left(\frac{T}{T^*_\textrm{u}}-\frac{T^*_\textrm{u}}{T}+2\ln\frac{T^*_\textrm{u}}{T}\right)\right],
\label{eq:psatfinal}
\end{align}
where $p^*_\textrm{sat,u}$ is the SVP at $T^*_\textrm{u}$.

\begin{figure}[h]
  \centering
    \includegraphics[width=\columnwidth]{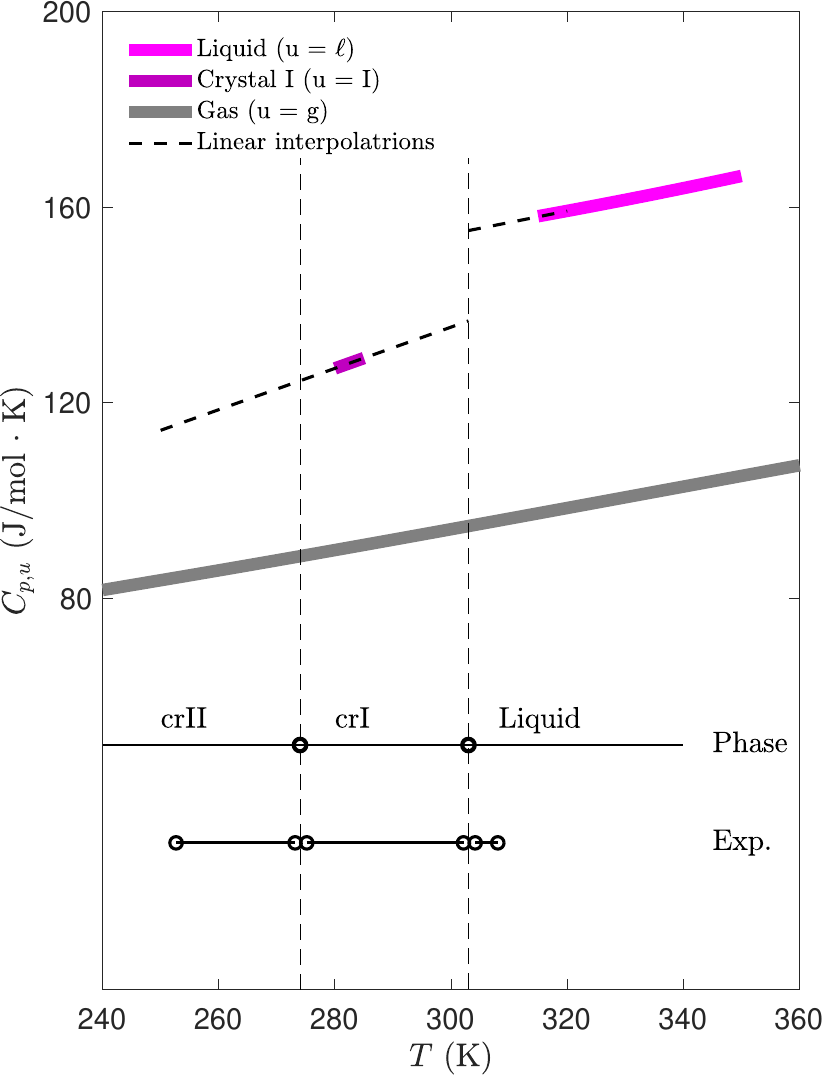}
    \caption{Variation of the heat capacities versus temperature. The pink, magenta and grey thick lines show the values of the liquid, solid (crI) and gaseous heat capacities at constant pressure reported in the literature. The black dashed lines show the results of the linear interpolations used to determine $\beta_\textrm{u}$ and $\alpha_\textrm{u}$ in the different temperature regions corresponding to the different phases. The vertical dashed lines show the positions of the phase transitions when increasing the temperature. The bottom horizontal black lines indicate the temperature ranges used in the experimental measurements 1 and 2.}
\label{fig:heat_capacities}
\end{figure}

The values of $\beta_\textrm{u}$ and $\alpha_\textrm{u}$ for the gaseous, liquid and crI phases can be obtained from linear interpolations of the values reported in the literature for the heat capacities at constant pressure~\cite{Stejfa2020}, as shown in Fig.~\ref{fig:heat_capacities}. In absence of reported values for the crII heat capacity we use values extrapolated from the crI phase data.

The determination of the SVP and enthalpies of the different phases from the measured chamber pressure $p_V$ as a function of sample temperature $T_s$ is performed on the basis of a statistical rate theory (SRT) model for the evaporation and condensation flux from the sample, as performed in~\cite{Nielsen2024,Salimi2025,Salimi2025b} for liquids. As long as the sample thermalization time is much slower than the evaporation and condensation rates, the net particle flux can be assumed to be zero (steady state condition) at any time/temperature, from which the chamber pressure can be related to the SVP via the relation
\begin{equation}
p_V=p_\textrm{sat,u}(T_S)\times f_\textrm{u}(T_S,T_V,\omega_i),
\end{equation}
where the characteristic function $f_\textrm{u}$ depends on the vibrational mode frequencies $\omega_i$ ($i=1,...,\textrm{DOF}$), where DOF=30 is the number of vibrational degrees of freedom of the molecule.

This characteristic function can be determined numerically using the steady state flux condition and approximating the vibrational degrees of freedom as harmonic in the SRT model~\cite{Nielsen2024}. However, as shown in~\cite{Salimi2025b}, it can be advantageous to introduce an effective number of vibrational degrees of freedom, $D_{e,\textrm{u}}$ ($0\leq D_{e,\textrm{u}}\leq\textrm{DOF}$), which can be determined from the knowledge of the gaseous heat capacity via
\begin{equation}
D_{e,\textrm{u}}=(C_{p,\textrm{g}}-4k_\textrm{B}N_\textrm{A})/(k_\textrm{B}N_\textrm{A})
\label{eq:De}
\end{equation}
and which hence takes into account anharmonicities and conformal effects. The pressure in the chamber can then be phenomenologically modelled by an analytical expression corresponding to the steady state result of the SRT model with an effective DOF $D_{e,u}$
\begin{equation}
p_V^{D_{e,\textrm{u}}}=p_\textrm{sat,u}(T_S)\times \exp\left[(D_{e,\textrm{u}}+4)\left(1-\frac{T_V}{T_S}\right)\right]\left(\frac{T_V}{T_S}\right)^{D_{e,\textrm{u}}+4}.
\label{eq:pVDe}
\end{equation}
The $p_V$ {\it vs} $T_S$ data can then be conveniently fitted in each temperature region with Eq.~(\ref{eq:pVDe}) using the previously determined values of $\alpha_\textrm{u}$, $\beta_\textrm{u}$ and $D_{e,\textrm{u}}$ and with $p_\textrm{sat,u}^*$ and $\Delta H_\textrm{u}^*$ as free fitting parameters.

The uncertainties of the derived quantities are dominated by the uncertainties of the temperature and pressure readouts. The sample temperature is determined by a self-assembled sensor (type K, chromel-alumel), calibrated against a commercially available and accurate PT100 sensor (SE02, Pico Technology) with a final estimated accuracy $\delta T_S=0.1^\circ$C. The specified pressure uncertainty for the pressure sensors is $\delta p=3\times 10^{-4}\textrm{Pa}+1.5\times 10^{-3}p_V$. The final uncertainties are reported with a $1\sigma$ deviation.

%%%%%%%%%%%%%%%%%%%%%%%%%%%

\section{Results}\label{sec:results}

\begin{figure}[h]
  \centering
    \includegraphics[width=\columnwidth]{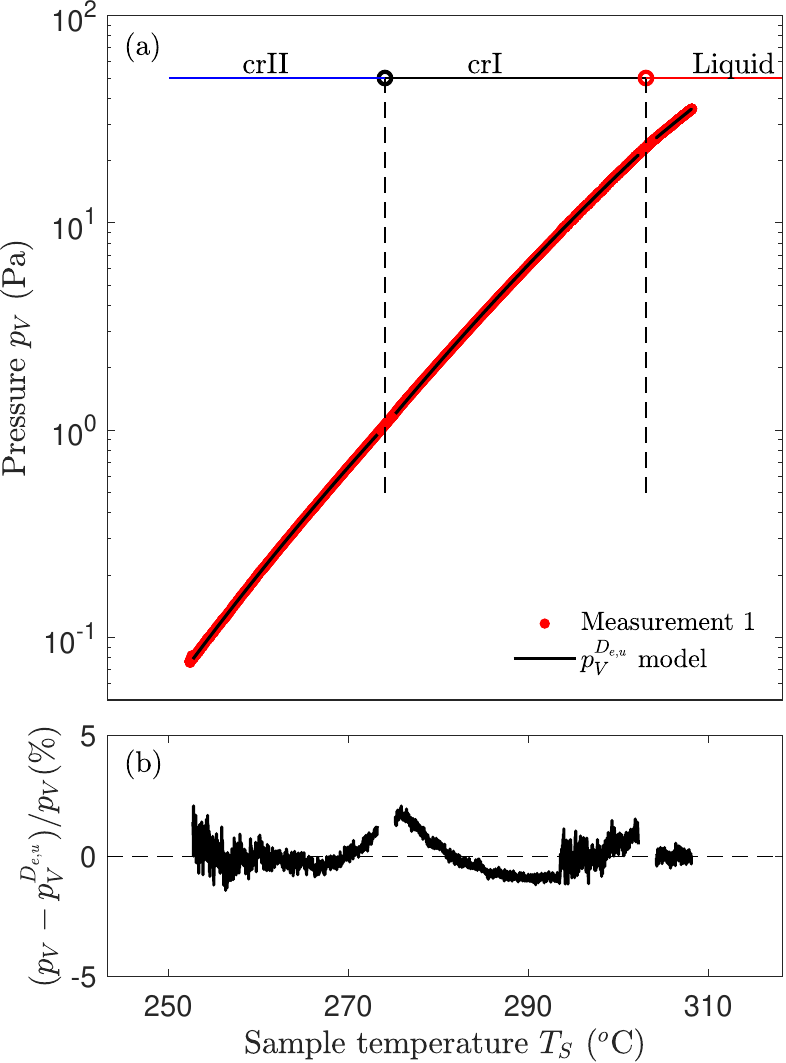}
    \caption{(a) $p_V$ {\it vs} $T_S$ for measurement 1 (red), together with the results of fits with Eq.~(\ref{eq:pVDe}) for each phase (thin black lines). (b) Fit residuals.}
\label{fig:results_pV_T}
\end{figure}

Figure~\ref{fig:results_pV_T} shows the variation of the chamber pressure $p_V$ with the sample temperature $T_S$ during measurement 1, the vertical dashed lines indicating the crI--crII and crII-liquid phase transitions for an increasing temperature. The thin black lines show the results of fits with the analytical model (Eq.~(\ref{eq:pVDe})) for each phase. As seen from the residuals, the model is observed to represent the data very well in each region. 

\begin{figure}[h]
  \centering
    \includegraphics[width=\columnwidth]{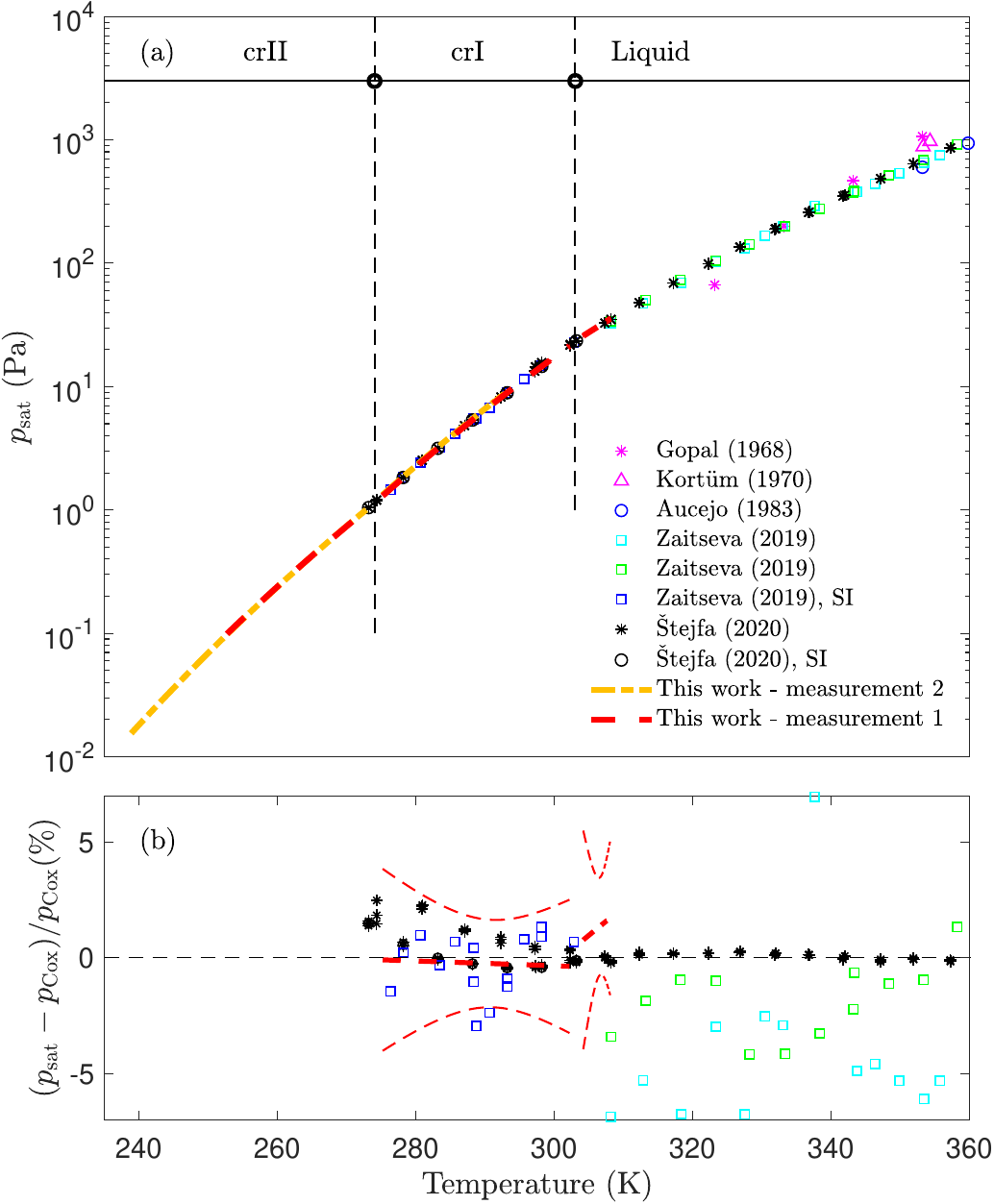}
    \caption{(a) $p_{\textrm{sat}}$ {\it vs} $T_s$ for measurements 1 (red line) and 2 (orange line), together with previously reported results~\cite{Gopal1968,Kortum1970,Aucejo1993,Zaitseva2019,Zaitseva2019b,Stejfa2020}. (b) Relative deviation of the measured SVPs with respect to the Cox parametrization performed in Ref.~\cite{Stejfa2020}. The thin red lines show the $\pm 1\sigma$ confidence interval for measurement 1. The symbols show the results of Refs.~\cite{Gopal1968,Kortum1970,Aucejo1993,Zaitseva2019,Zaitseva2019b,Stejfa2020}.}
\label{fig:results_ps}
\end{figure}

\begin{figure}[h]
	\centering
	\includegraphics[width=\columnwidth]{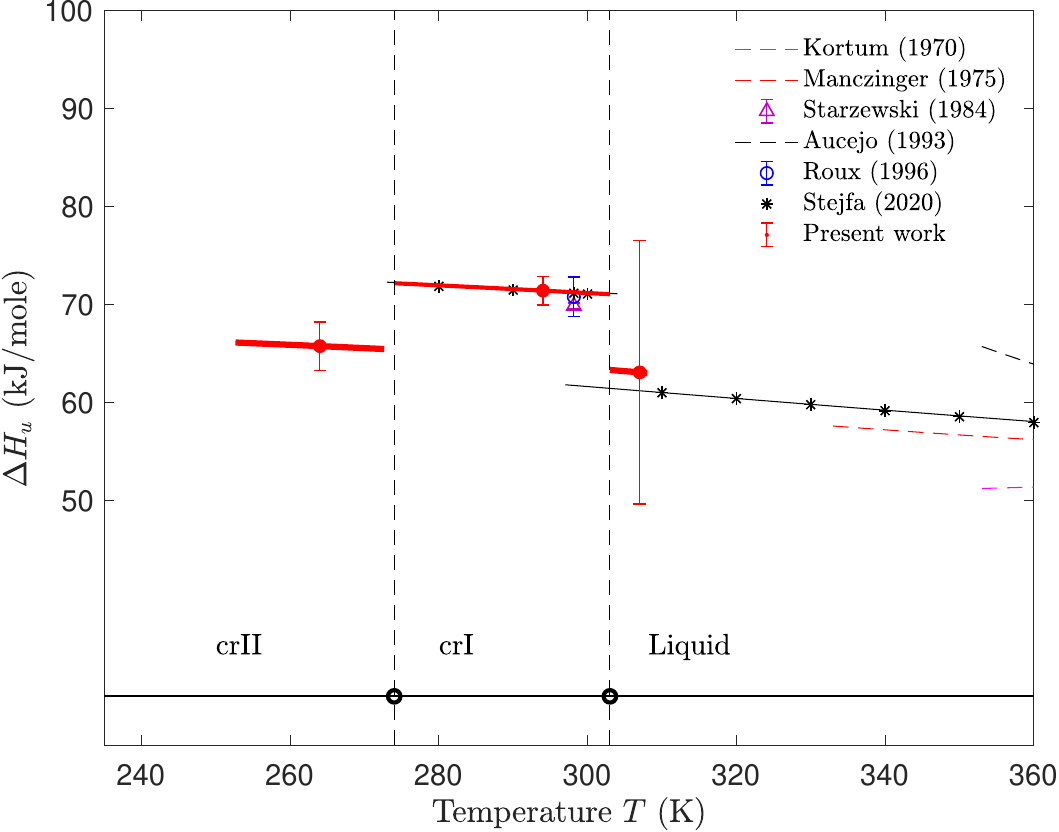}
	\caption{Enthalpies of sublimation and vaporization for $N$-methyl acetamide determined in this work for the three phases as a function of temperature. The single points with error bar indicate the used reference temperature ($T^*_\textrm{u}$) and show the $1\sigma$ confidence interval. The black dashed line shows the Cox parametrization of Ref.~\cite{Stejfa2020}. The symbols show the results of Refs.~\cite{Kortum1970,Manczinger1975,Starzewski1984,Aucejo1993,Roux1996,Stejfa2020}, while the dashed lines indicate the results of Refs.~\cite{Kortum1970,Manczinger1975,Aucejo1993}, which are given as the Antoine parameterizations performed in Ref.~\cite{MacKerell2008}.
}
\label{fig:results_DH}
\end{figure}

The values for the SVP and enthalpies in each temperature range are shown as the dashed red and orange lines in Fig.~\ref{fig:results_ps}(a) and Fig.~\ref{fig:results_DH}. These values (and their uncertainties) can be readily computed via Eq.~(\ref{eq:psatfinal}) from the values of $p^*_\textrm{sat,u}$ and $\Delta H^*_\textrm{u}$ at $T^*_\textrm{u}$, which are reported in Table~\ref{tab:results} together with the values of $\beta_\textrm{u}$ and $\alpha_\textrm{u}$. The good consistency between the SVP and enthalpy values for the solid phases obtained in the two measurements with different chamber temperatures also indicates that only one phase is present at any time during the measurement, which would not be the case if, for instance, inhomogeneous melting or multi-crystalline structures occured progressively rather than abrupty at the phase transitions.

We also observe good agreement between the SVPs and enthalpies for both crI and liquid phase determined in this work and those reported previously~\cite{Gopal1968,Kortum1970,Aucejo1993,Zaitseva2019,Zaitseva2019b,Stejfa2020}. The relative deviation with the Cox parametrization for the SVP reported in Ref.~\cite{Stejfa2020} are shown in Fig.~\ref{fig:results_ps}(b). We also report for the first time values for the saturation vapor pressure and enthalpy of sublimation of $N$-methyl acetamide crII in the range $-30$-$0^\circ$C. We observe a significantly lower value for the enthalpy of sublimation for the crII phase than for the more disordered crI phase.

\begin{table*}[h]
\caption{Summary of experimental conditions, estimated temperature variations for the enthalpies of sublimation/evaporization ($T^*_\textrm{u}$, $\beta_\textrm{u}$, $\alpha_\textrm{u}$, see Eq.~(\ref{eq:DH})), effective number degrees of freedom ($D_{e,\textrm{u}}$, see Eq.~(\ref{eq:De})), as well as the results of model fits (Eq.~(\ref{eq:pVDe})) giving SVP ($p^*_\textrm{sat,u}$) and enthalpies ($\Delta H^*_\textrm{u}$) for $N$-methyl-acetamide. The fit results are given with a $1\sigma$ (63\%) confidence interval.}
\label{tab:results}
\begin{tabular*}{\tblwidth}{@{}LL@{}LL@{}LL@{}LL@{}LL@{}LL@{}LL@{}LL@{}}
\toprule
  & Phase & Range & $T^*_u$ (K) & $\beta_u$ (kJ/mol) & $\alpha_u$ (kJ/mol/K) & $D_{e,u}$ & $p^*_\textrm{sat,u}$ (Pa) & $\Delta H^*_u$  (kJ/mol) \\
% & & (K) & (kJ/mol) & (kJ/mol/K) & & (Pa) & (kJ/mol) \\% Table header row
\midrule
Measurement 1 		& crI & $-20$-$0^\circ$C 		& 264 & $-33.8$ & $-0.21$ & 6.4 & $0.38\pm0.01$ & $65.8\pm2.5$\\
$T_V=36.4^\circ$C	& crII & $2$-$29^\circ$C 		& 294 & $-39.9$ & $-0.21$ & 7.0 & $9.7\pm0.2$ 	& $71.4\pm1.4$\\
					 & Liquid & $31$-$34.5^\circ$C & 307 & $-60.4$ & $-0.02$ & 7.5 & $32.5\pm0.7$ & $63.1\pm13.5$\\
\midrule
Measurement 2 		& crI & $-34$-$0^\circ$C 		& 264 & $-33.8$ & $-0.21$ & 6.2 & $0.38\pm0.01$ & $66.3\pm0.9$\\
$T_V=20.5^\circ$C	& crII& $2$-$19.5^\circ$C 	& 294 & $-39.9$ & $-0.21$ & 6.9 & $9.9\pm0.2$ & $71.6\pm1.6$\\
\bottomrule
\end{tabular*}
\end{table*}

%%%%%%%%%%%%%%%%%%%%%%%%%%%%%%%%%%%%%%%%

\section{Conclusion}\label{sec:conclusion}

We reported on the determination of the SVP of $N$-methyl acetamide in the temperature range $-30^\circ$C to $34^\circ$C and determined, in particular, for the first time the SVP and enthalpy of sublimation of crII NMA in the range $-30$-$0^\circ$C. In addition to providing new thermodynamical data for NMA this work demonstrates that it is possible to determine---in a single run---the SVP of a low-volatide substance over a broad temperature range, as it successively undergoes solid-solid and solid-liquid phase transitions. When combined with low pressure measurement capabilities as demonstrated in, e.g.,~\cite{Salimi2025b,Salimi2024}, we expect the method to be applicable to a wide range of other polymorphic low-volatile substances, which will be the subject of future investigations.


\printcredits

\bibliographystyle{cas-model2-names}

\bibliography{Nmethylacetamide_bib}

@article{Katz1960,
author = "Katz, J. L. and Post, B.",
title = "{The crystal structure and polymorphism of {\it N}-methylacetamide}",
journal = "Acta Crystallographica",
year = "1960",
volume = "13",
number = "8",
pages = "624--628",
month = "Aug",
doi = {10.1107/S0365110X60001485},
url = {https://doi.org/10.1107/S0365110X60001485},
}

@article{Halfen2011,
doi = {10.1088/0004-637X/743/1/60},
url = {https://doi.org/10.1088/0004-637X/743/1/60},
year = {2011},
month = {nov},
publisher = {The American Astronomical Society},
volume = {743},
number = {1},
pages = {60},
author = {Halfen, D. T. and Ilyushin, V. and Ziurys, L. M.},
title = {FORMATION OF PEPTIDE BONDS IN SPACE: A COMPREHENSIVE STUDY OF FORMAMIDE AND ACETAMIDE IN Sgr B2(N)},
journal = {The Astrophysical Journal}
}

@Book{Kirk1991,
author={Kirk, R. E. and Othmer, D. F. and Kroschwitz, J. I. and Howe-Grant, M.},
title={Kirk-Othmer encyclopedia of chemical technology, vol. 1},
publisher={Wiley, New York, NY},
year={1991},
}

@article{Bernauer2008,
author = {Bernauer, Milan and Dohnal, Vladimir},
title = {Temperature Dependence of Air-Water Partitioning of N-Methylated (C1 and C2) Fatty Acid Amides},
journal = {Journal of Chemical and Engineering Data},
volume = {53},
number = {11},
pages = {2622-2631},
year = {2008},
doi = {10.1021/je800517r},
URL = {https://doi.org/10.1021/je800517r},
eprint = {https://doi.org/10.1021/je800517r}
}

@article{Ashford1995,
title = {Ashford's Dictionary of Industrial Chemicals},
journal = {Analytical Chemistry},
volume = {67},
number = {11},
pages = {385A-385A},
year = {1995},
doi = {10.1021/ac00107a730},
note ={PMID: 22853681},
URL = {https://doi.org/10.1021/ac00107a730},
eprint = {https://doi.org/10.1021/ac00107a730}
}

@article{Stejfa2020,
title = {Thermodynamic study of acetamides},
journal = {Journal of Molecular Liquids},
volume = {319},
pages = {114019},
year = {2020},
issn = {0167-7322},
doi = {https://doi.org/10.1016/j.molliq.2020.114019},
url = {https://www.sciencedirect.com/science/article/pii/S0167732220339921},
author = {Vojtěch Štejfa and Sothys Chun and Václav Pokorný and Michal Fulem and Květoslav Růžička}
}

@article{Zaitseva2019,
title = {Vapour pressures and enthalpies of vaporisation of N‑alkyl acetamides},
journal = {Journal of Molecular Liquids},
volume = {293},
pages = {111453},
year = {2019},
issn = {0167-7322},
doi = {https://doi.org/10.1016/j.molliq.2019.111453},
url = {https://www.sciencedirect.com/science/article/pii/S0167732219320835},
author = {Ksenia V. Zaitseva and Mikhail A. Varfolomeev and Sergey P. Verevkin}
}

@article{Zaitseva2019b,
author={Zaitseva, K. V. and Varfolomeev, M. A: and Verevkin, S. P.},
title={Vapour pressures and enthalpies of vaporization of N,N-di-alkyl-acetamides},
journal={Fluid Phase Equilibria},
volume={499},
pages={112241},
year={2019}
}

@article{Aucejo1993,
author = {Aucejo, Antonio and Monton, Juan B. and Munoz, Rosa and Sanchotello, Margarita},
title = {Isobaric vapor-liquid equilibrium data for the cyclohexanone + N-methylacetamide system},
journal = {Journal of Chemical \& Engineering Data},
volume = {38},
number = {1},
pages = {160-162},
year = {1993},
doi = {10.1021/je00009a039},
URL = {https://doi.org/10.1021/je00009a039},
eprint = {https://doi.org/10.1021/je00009a039}
}

@article{MacKerell2008,
author = {MacKerell, Alexander D. Jr. and Shim, Ji Hyun and Anisimov, Victor M.},
title = {Re-Evaluation of the Reported Experimental Values of the Heat of Vaporization of N-Methylacetamide},
journal = {Journal of Chemical Theory and Computation},
volume = {4},
number = {8},
pages = {1307-1312},
year = {2008},
doi = {10.1021/ct8000969},
URL = {https://doi.org/10.1021/ct8000969},
eprint = {https://doi.org/10.1021/ct8000969}
}

@article{Gopal1968,
author={Gopal, Ram and Rizvi, Sharaf Abbas},
title={Vapour Pressures of some Mon- and Di-Alkyl Substituted Aliphatic Amides at Different Temperatures},
journal={Journal of Indian Chemical Society},
volume={45},
pages={13-16},
year={1968}
}

@article{Kortum1970,
author={Kort\"{u}m, G. and Biedersee, H. v.},
title={Dampf/Fl\"{u}ssigkeit-Gleichgewichte (Siedediagramme) bin\"{a}rer Systeme hoher relativer Fl\"{u}chtigkeit},
journal={Chemie-Ing.-Tech.},
volume={42},
pages={552-560},
year={1970}
}

@article{Roux1996,
author = {Roux, M. V. and Jim{\'e}nez, P. and Dávalos, J. Z. and Casta{\~n}o, O. and Molina, M. T. and Notario, R. and Herreros, M. and Abboud, J.-L. M.},
title = {The First Direct Experimental Determination of Strain in Neutral and Protonated 2-Azetidinone},
journal = {Journal of the American Chemical Society},
volume = {118},
number = {50},
pages = {12735-12737},
year = {1996},
doi = {10.1021/ja962792w},
URL = {https://doi.org/10.1021/ja962792w},
eprint = {https://doi.org/10.1021/ja962792w}
}

@article{Manczinger1975,
author={Manczinger, J. and Kort\"{u}m, G.},
title={Thermodynamische Mischungseffekte im System Wasser(1)/N-Methylacetamid(2)},
journal={Zeitschrift f\"{u}r Physikalische Chemie Neue Folge},
volume={95},
pages={177-186},
year={1975}
}

@article{Starzewski1984,
author={Starzewski, P. and Wads\"{o}, I. and Zielenkieewicz, W.},
title={Enthalpies of vaporization of some N-alkylamides at 298.15 K},
journal={J. Chem. Thermodynamics},
volume={16},
pages={331-334},
year={1984}
}

@Article{Clarke1966,
author ="Clarke, E. C. W. and Glew, D. N.",
title  ="Evaluation of thermodynamic functions from equilibrium constants",
journal  ="Trans. Faraday Soc. ",
year  ="1966",
volume  ="62",
issue  ="0",
pages  ="539-547",
publisher  ="The Royal Society of Chemistry",
doi  ="10.1039/TF9666200539",
url  ="http://dx.doi.org/10.1039/TF9666200539"}

@Book{wark1988,
author={Wark, Kenneth},
title={"Generalized Thermodynamic Relationships." Thermodynamics (5th ed.)},
publisher={New York, NY: McGraw-Hill, Inc.},
year={1988}
}

@Article{Nielsen2024,
author={Nielsen, Robin V. and Salimi, M. and Andersen, John V. and Elm, Jonas and Dantan, Aurelien and Pedersen, Henrik B.},
title={A new setup for measurements of absolute saturation vapor pressures using a dynamical method: Experimental concept and validation},
journal={Rev. Sci. Instr.},
year={2024},
volume={95},
pages={065007},
url={https://doi.org/10.1063/5.0215176},
}

@Article{Salimi2025,
author={Salimi, M. and Nielsen, Robin V. and Elm, Jonas and Dantan, Aurelien and Pedersen, Henrik B.},
title={Absolute Saturation Vapor Pressures of Three Fatty Acid Methyl Esters around Room Temperature},
journal={ACS Omega},
year={2025},
volume={10},
pages={6671-6678},
url={https://doi.org/10.1021/acsomega.4c08095},
}

@Article{Salimi2025b,
author={Salimi, M. and Pedersen, Andreas B. and Andersen, J. E. V. and  Pedersen, Henrik B. and Dantan, Aurelien},
title={Dynamical measurement of saturation vapor pressures below and above room temperature},
journal={arXiv:2508.10932},
year={2025},
volume={},
pages={},
url={},
}

@Article{salimi2024,
  author   = {Mohsen Salimi and Robin V. Nielsen and Henrik B. Pedersen and Aurélien Dantan},
  title    = {Squeeze film absolute pressure sensors with sub-millipascal sensitivity},
  journal  = {Sensors and Actuators A: Physical},
  year     = {2024},
  volume   = {374},
  pages    = {115450},
  issn     = {0924-4247},
  doi      = {https://doi.org/10.1016/j.sna.2024.115450},
  url      = {https://www.sciencedirect.com/science/article/pii/S0924424724004448},
}

\end{document}